
\magnification=\magstep1
\hsize=5.9truein
\vsize=8.375truein
\parindent=25pt
\nopagenumbers
\raggedbottom
\hangafter=1
\def\makeheadline{\vbox to 0pt{\vskip-40pt
   \line{\vbox to 8.5pt{}\the\headline}\vss}\nointerlineskip}
\def\approxlt{\kern 0.35em\raise 0.45ex\hbox{$<$}\kern-0.66em\lower0.5ex
   \hbox{$\scriptstyle\sim$}\kern0.35em}
\def\approxgt{\kern 0.35em\raise 0.45ex\hbox{$>$}\kern-0.75em\lower0.5ex
   \hbox{$\scriptstyle\sim$}\kern0.35em}
\vglue1.5truein
\centerline{\bf A 2D Inspired 4D Theory of Gravity}
\vskip22pt
\centerline{ V.G.J. Rodgers}
\centerline{ Department of Physics and Astronomy}
\centerline{ The University of Iowa}
\centerline{ Iowa City, Iowa~~52242--1479}
\centerline{ July, 1994 }
\vglue1.0truein
\baselineskip=12pt
\centerline{\bf ABSTRACT}
\vskip22pt
Coadjoint orbits of the Virasoro and Kac-Moody algebras provide geometric
actions for matter coupled to gravity and gauge fields in two dimensions.
However, the Gauss' law constraints that arise from these actions are not
necessarily endemic to two-dimensional topologies.
Indeed the constraints associated with Yang-Mills naturally arise from
the coadjoint orbit construction of the WZW model.  One may in fact use
a Yang-Mills theory to provide dynamics to the
otherwise fixed coadjoint vectors that define the orbits.  In this letter we
would like to exhibit an analogue of the Yang-Mills classical action for
the diffeomorphism sector.  With this analogue one may postulate a 4D
theory of gravitation that is related to an underlying two dimensional
theory.  Instead of  quadratic differentials, a (1,3) pseudo tensor becomes
the dynamical variable.  We briefly discuss how this tensor may be classically
coupled to matter.
\vfill
\eject
\headline={\tenrm\hfil\folio}
\baselineskip=16pt
\pageno=1
Coadjoint orbits have enjoyed much success in the explanation of
fermions coupled to gauge fields and  quantum gravity in two dimensions,
see for example Ref. [1-7].  Furthermore two dimensions have
provided a platform for many insights into four-dimensional theories.
It would appear by their very construction
that the geometric
actions associated with coadjoint orbits are necessarily two-dimensional
theories and provide little insight into theories of higher dimension, in
particular four dimensions.
However, the isotropy equations for the orbits yield Gauss' law constraints
that are directly related to the underlying gauge and coordinate
transformations in the theory.  If we are willing to dispense with conformal
invariance
of the Riemann surface, these constraints can be used to build theories in
higher dimensions.  For example, in the Kac-Moody sector, the isotropy
condition on the orbit can be interpreted as the Gauss' law constraint
from Yang-Mills theories.
Using this one can provide dynamics to coadjoint vectors by introducing a
Yang-Mills action.  A similar construction has been attempted
in 2D for the diffeomorphism sector [8,9].  In this note
we take the position that the Gauss' law constraints that arise from the
coadjoint orbit construction are not endemic to two dimensions but are
dimensionally reduced constraints from a dimensionally independent theory
such as Yang-Mills. With this one may
construct an action for gravity in 4D that exhibits the transformation
laws and constraints of its two-dimensional partner.

To proceed, let us recapitulate the salient features of the coadjoint orbit
construction of fermions coupled to matter in 2D.  There one begins with
the centrally extended algebra corresponding to 2D gauge and
coordinate transformations,
$$
\eqalign{
 [ J^{\alpha}_N, J^{\beta}_M] & = i f^{\alpha\beta\gamma} J^\gamma_{N+M} +
N k \delta_{M+N,0} \delta^{\alpha\beta} \cr
[ L_N,J^{\alpha}_M] & = - M J^{\alpha}_{M+N} \cr
[ L_N,L_M ] &= (N-M) L_{N+M} + {c\over 12} (N^3-N) \delta_{N+M,0},}
$$
where we are using notation of ref. [8].  The $L's$ and the
$J's$ provide a generic basis for the adjoint representation, i.e.
$F_{AB}^{\beta}(\rho) = (L_A,J^\beta_B,\rho)$, or more  generally
$F = (\xi (\theta), \Lambda(\theta), a)$.  The dual elements of the
adjoint representation are denoted as $B(\theta) = (T(\theta),
A_\theta(\theta),\mu )$ and can be used to define the coadjoint representation
through
$$\eqalign{
&\delta_F B \equiv (\xi(\theta),\Lambda(\theta),a) \ast
(T(\theta),A_\theta(\theta),
\mu) = \cr
-&\bigl( 2 \xi'T+T'\xi + {c \mu\over 24\pi} \xi''' - Tr[ A_\theta \Lambda'],
 A_\theta'\xi+A_\theta\xi' + [\Lambda A_\theta - A_\theta \Lambda] +  k \mu
\Lambda', 0 \bigr)\cr
= & (\delta T, \delta A_\theta, 0).}\eqno(1)
$$
The parameters $T$ and $A_\theta$ are the residual fields left to couple to
the fermions after gauge fixing.  From the first component in the
transformation laws of $B$, we see
that $T$ is a rank two {\it pseudo} tensor. This is due to the
inhomogeneous  term $\xi'''$.  Also, observe that this pseudo tensor can have
$n$ contravariant
indices and $n+2$ covariant indices and still appear to transform like
a rank two pseudo tensor when gauge fixed to a circle.  Another remarkable
feature of $T$ is that it
transforms under ${\it gauge}$ transformations via Tr$[\Lambda' A_\theta]$.
Since we have taken the position that these transformation laws
can be extended to dimensions other than two we cannot dismiss this
gauge transformation as a two-dimensional artifact.  This  suggests that
in a theory of gravity with matter, the idea of extracting a pure
``graviton'' propagator may not make sense.  The ``background'' gauge
fields must be present in the graviton propagator just as the metric
is present in the propagator for the vector potentials.  Throughout this
note we will put these issues aside and discuss them in a larger work.
As far as $A_\theta$ is concerned
one can quickly identify $A_\theta$ as the Yang-Mills vector potential.
Although we claim that the transformation laws can be extended to
dimension four (say), we do not expect the parameter ${c/( 24\pi)}$
to be constrained to the dimension of the group and the dual coxeter
number of the representation.  A similar constraint may hold for four
dimensions but this is
unknown.  Therefore from hereon we will refer to the parameter in front
of the inhomogeneous term simply as ${1/ q}$.

{}From the isotropy equations of the orbits ($\delta B = 0$) we infer what
the Gauss' law is for our system.  Setting $\delta B = 0$ we have
$$ 2 f' T + f T' + {1\over q} f''' = 0. \eqno(2) $$   Since the
two-dimensional theory dictates that $T$ couples to the induced metric of
fermions in the geometric action via $ \int T h_{\tau \tau} d^2 \sigma$, we
deduce that $T$ is
equivalent to the $T_{\theta \theta}$ component of a rank two pseudo tensor.
Then, keeping track of the local Lorentz invariance in the two dimensional
system, the Gauss' law above is of a form like
$$
\partial_\theta T_{\theta \theta} \partial_\tau T_{ \theta \tau}
+ 2 T_{\theta \theta} \partial_\tau \partial_\theta T_{ \theta \tau}
+ c {\partial_\theta}^3 \partial_\tau T_{ \theta \tau} = 0, \eqno(3)
$$
where the gauge fixing conditions has set $T_{ \tau \tau} = 0$.
Therefore a reasonable set of conditions for our action is that

a) it must admit a rank two (or $n$ contravariant and $n+2$ covariant indices)
pseudo
tensor,

b) the 2D gauge fixed reduction of the coordinate transformation
for $T$ must correspond to Eq.~(2),

c) it have no dynamics for some components of $T$,

d) there may be no higher than fourth order derivatives in the action, and

e) it be generally covariant.

The two dimensional transformation laws also suggest that $T$ has mass
dimension two, since ${c/ (2 \pi)}$ is assumed to be dimensionless
number. However, if needed one may use the gravitational constant, $\kappa$,
to construct the dimensionless field  $\kappa T$.  We will return to this
when we discuss coupling this field to matter.  The origin of the inhomogeneous
term may be quantum field theoretic, however, there is no apriori reason to
assume this.  In fact, the two-dimensional case would put the inhomogeneous
term on the same footing as the gauge potential's inhomogeneous term.
Also, since $T$ couples to the induced metric of the two dimensional theory
we may think of $T$ as a modified ``stress tensor for gravity'' [10,11].
With this in place we can construct a suitable $T$ field.

For $g= -1$ (the determinant of the metric) one may construct a
locally conserved stress pseudo tensor for the background
metric as [10],
$$ \kappa t^\alpha_\sigma = {1 \over 2} \delta^\alpha_\sigma g^{\mu \nu}
\Gamma^\lambda_{ \mu \beta} \Gamma^\beta_{\nu \lambda} - g^{\mu \nu}
\Gamma^\alpha_{\mu \beta} \Gamma^\beta_{\nu \sigma}. \eqno(4)$$
Although this is a pseudo tensor, it does not transform as $\xi'''$
since there is never more than one derivative at any instance on $g_{\mu \nu}$.
But the
Riemann curvature tensor can provide a clue as to the nature of $T$.
Write $R^\mu_{\sigma \alpha \beta}$ as
$$
R^\mu_{\sigma \alpha \beta} = \nabla_\alpha \Gamma^\mu_{\sigma \beta}
- \nabla_\beta \Gamma^\mu_{\sigma \alpha} + \Gamma^\nu_{\alpha \sigma}
\Gamma^\mu_{\nu \beta} - \Gamma^\mu_{\alpha \nu}
\Gamma^\nu_{\sigma \beta}. \eqno(5) $$
Define a rank $(1,3)$ pseudo tensor as the ``background stress energy
potential'' through
$$\Theta ^\mu_{\sigma \alpha \beta} = \Gamma^\mu_{\alpha \nu}
\Gamma^\nu_{\sigma \beta}. \eqno(6)$$
There are two geometrically significant contractions that one can form on this
``potential'' namely
a Ricci and a Weyl contraction, where respectively these are,
$$
\Theta^R_{\alpha \beta} = g^{\mu \nu} g_{\alpha \lambda} \Theta^\lambda_{\nu
\mu \beta} \eqno(7) $$ and
$$
\Theta^W_{\alpha \beta} = g^{\sigma \nu} g_{\alpha \lambda} \Theta^\lambda_{\nu
\beta \sigma}. \eqno(8) $$
{}From $\Theta^R_{\alpha \beta}$ one can recover the gravitational stress
pseudo tensor
of Einstein's as
$$
t^R_{\alpha \beta} = {1 \over 2} g_{\alpha \beta} \Theta^R - \Theta^R_{\alpha
\beta}, \eqno(9) $$ where here $g = -1$ is assumed.  The Weyl contraction to
the
stress tensor for gravity can be written as
$$ t^W_{\alpha \beta} = {1 \over 2} g_{\alpha \beta} \Theta^W -
\Theta^W_{\alpha \beta}, \eqno(10)$$ and will vanish if $g = -1$.  Keep in mind
that even when
these pseudo tensors are zero (pure gauge), they still transform
inhomogeneously.

Now let us focus on the pseudo tensor $\nabla_\alpha \Gamma^\mu_{\sigma \beta}$
that is also present in the Riemann curvature tensor.  In the 2D gauge fixed
limit this does transform as $\xi '''$.  With this we define our tensor
$T$ as the rank (1,3) tensor $T^\mu_{\alpha \sigma \beta}$ that transforms
like ${1/ q} \nabla_\alpha \Gamma^\mu_{\sigma \beta}$, where $q$ is a
dimensionless constant.  In other words
$$
\delta T^\mu_{\alpha \sigma \beta} = (T^\mu_{\alpha \sigma \beta})_{\rm tensor}
+ {1\over q} \nabla_\alpha \partial_\sigma \partial_\beta \xi^\mu . \eqno(11)
$$
One may also consider Ricci and Weyl contractions with it to form,
$$
T^R_{\alpha \beta} = g^{\mu \nu} g_{\alpha \lambda} T^\lambda_{\nu \mu \beta}
\eqno(12) $$ and
$$
T^W_{\alpha \beta} = g^{\sigma \nu} g_{\alpha \lambda} T^\lambda_{\nu \beta
\sigma}. \eqno(13) $$

 Now that we have a
suitable pseudo tensor we can construct an action with it.

Define a tensor (not pseudo) $K^\mu_{\lambda \rho \alpha \sigma \beta}$ as
$$\eqalign{
K^\mu_{\lambda \rho \alpha \sigma \beta} =&
{}~q \nabla_\lambda \nabla_\rho T^\mu_{\alpha \sigma \beta}
- q \nabla_\lambda \nabla_\rho T^\mu_{\beta \sigma \alpha }\cr
&+q^2 T^\nu_{\rho \alpha \sigma} T^\mu_{\lambda \nu \beta}
+q^2 T^\nu_{\lambda \alpha \sigma} T^\mu_{\rho \nu \beta}
-q^2 T^\mu_{\rho \alpha \nu } T^\nu_{\lambda \sigma \beta}
-q^2 T^\mu_{\lambda \alpha\nu } T^\nu_{\rho \sigma \beta} \cr
& + q \nabla_\lambda T^\nu_{\rho \alpha \nu} \Gamma^\mu_{\nu \beta}
+ q \nabla_\lambda T^\mu_{\rho \nu \beta} \Gamma^\nu_{\alpha \sigma}
-q \nabla_\lambda T^\mu_{\rho \alpha \nu} \Gamma^\nu_{\sigma \beta}
-q \nabla_\lambda T^\nu_{\rho \sigma \beta } \Gamma^\mu_{\alpha
\nu},}\eqno(14)$$
where we have exhibited the coupling of the background $\Gamma$ to $T$
which is necessary for general covariance.  In order to be consistent with
the 2D theory recall we must keep in mind that the Gauss' law constraint came
from a term in the action like,
$$
T_{\tau \tau} ( \partial_\theta T_{\theta \theta} \partial_\tau T_{\theta \tau}
+ 2 T_{\theta \theta} \partial_\theta \partial_\tau T_{\theta \tau} + c~
\partial_\theta^3 \partial_\tau T_{\theta \tau}).$$  This may be rewritten
as
$$
T_{\theta \theta} T_{\tau \tau}\partial_\theta \partial_\tau T_{\theta \tau}
+ c~ \partial_\theta^2 T_{\tau \tau} \partial_\theta \partial_\tau T_{\theta
\tau}, $$
and tells us that the Lagrangian is the square of a tensor which has
the structure, $ c ~\partial \partial T + T T $, and that the derivative
operators are contracted opposite to the indices of  $T$.
With this, we write our invariant action which corresponds to the
2D action as
$$
S = \kappa^2 \int \sqrt g~ g^{\mu \gamma} g^{\alpha \sigma} K_{\mu \gamma}
 K_{\alpha \sigma}   ~d^4x, \eqno(15)
$$
where $\kappa$ is the gravitational constant,
$$
g^{\lambda \gamma \rho \beta} = g^{\lambda \beta} g^{\gamma \rho} -
g^{\lambda \rho } g^{\gamma \beta},~~ {\rm and}~~ K_{\alpha \sigma} =
g_{\mu \gamma} g^{\lambda \gamma \rho \beta} K^\mu_{\lambda \rho \alpha \sigma
\beta}. \eqno(16)
$$ One can see from the $\nabla_\rho \nabla_\lambda$ terms in  $K$ that only
the $(T^R)_{\alpha \beta}$  and $ (T^W)_{\alpha \beta}$ parts of $T$
propagate,
$$
g^{\lambda \gamma \rho \beta}g^{\lambda' \gamma' \rho' \beta'}(T^R_{\gamma
\beta} - T^W_{\gamma \beta}) \nabla_\lambda \nabla_\rho \nabla_{\lambda'}
\nabla_{\rho'}(T^R_{\gamma \beta} - T^W_{\gamma \beta}).
$$
  Furthermore a 2D reduction will lead to a Lagrangian
with terms like
$$
(-q \nabla_\lambda \nabla_\rho (T^W)^\mu_\beta g_{\mu \gamma} g^{\alpha \sigma}
g^{\lambda \gamma \rho \beta} + (T^W)^\nu_\beta (T^R)^\rho_\nu +
 q \nabla_\lambda \nabla_\rho (T^R)^\mu_\beta g_{\mu \gamma} g^{\alpha \sigma}
g^{\lambda \gamma \rho \beta})^2.
$$
Setting the $(T^R)_{\tau \tau}$ to zero as a gauge fixing condition yields the
analogous two-dimensional term.
In this theory, the propagator goes like $p^{-4}$ which is a good
 sign for convergent diagrams.  However, we must still check for unitarity,
the possibility of ghosts, and whether there is any acausal contributions  in
this action.  This will be studied in a longer work.

In order to show how $T$ may couple to matter, recall that in the
2D picture $T$ couples to the induced metric.
This induced metric is a bosonization of a fermion bilinear and
can be thought of as proportional to the fermion's stress energy tensor.
We can construct the analogous coupling with care that it be general coordinate
invariant.  Let $T^{\rm ~matter}_{\mu \nu} $ be the stress energy tensor
that is coupled to the background (classical) metric $g_{\mu \nu}$.
Then we can write the matter coupling as
$$
S_{\rm ~matter} = \int d^4 x \sqrt{g} ( (T^R)^{\mu \beta} - (T^W)^{\mu \beta}
+ (\Theta^W)^{\mu \beta} - (\Theta^R)^{\mu \beta}) T^{\rm ~matter}_{\mu \beta}.
\eqno(17)
$$
Note that this suggests that the background metric $g_{\mu \nu}$ is
modified to form the dynamical metric
$$
G_{\mu \nu} = g_{\mu \nu} +   \kappa ( (T^R)_{\mu \nu} - (T^W)_{\mu \nu}
+ (\Theta^W)_{\mu \nu} - (\Theta^R)_{\mu \nu}).
$$
The background stress tensors have been incorporated to  assure that
$G_{\mu \nu} $ is still a tensor.

Finally, there is the question of the gauge transformations that we
mentioned earlier.  Consider the pseudo tensor
$$
{\hat T}^\alpha_{\beta \rho \mu} = T^\alpha_{\beta \rho \mu} + {\rm
Tr}[g^{\alpha \lambda} A_\lambda A_\beta g_{\rho \mu}].
$$ Since the trace term is a real tensor, ${\hat T}$ still transforms as
a pseudo tensor.  However, this pseudo tensor is gauge invariant,
and can be used in the presence of gauge fields.

In conclusion, we have exhibited an action for gravity in 4D
that yields the same type of
Gauss' Law constraints as gravity in two dimensions.  This is analogous
to the origins of the Yang-Mills action in two dimensions.
The propagating field $T^\alpha_{\beta \gamma \rho}$ couples to a background
metric and carries the local diffeomorphism  information to matter.  The
usual metric provides a fixed background field, but one can write a
dynamical metric in terms of contractions of $T^\alpha_{\beta \gamma \rho}$ and
the background energy-momentum tensors.

\vfill\eject
\nopagenumbers
\baselineskip=13pt
\centerline{\bf REFERENCES}
\settabs 1 \columns
\+ [1] B. Rai and V.G.J. Rodgers, Nucl. Phys. B341(1990) 119 \cr
\+ [2] A. Yu. Alekseev and S.L. Shatashvili,  Mod. Phys. Lett. A3 (1988) 1551
\cr
\+ [3] P.B. Wiegmann, Nucl. Phys. B323 (1989) 311 \cr
\+ [4] A.M. Polyakov, Mod. Phys. Lett. A2 (1987) 893 \cr
\+ [5] A. Pressley and G. Segal, {\it Loop Groups}, (Oxford Univ. Press,
Oxford, 1986) \cr
\+ [6] E. Witten, Comm. Math. Phys. 114 (1988) 1 \cr
\+ [7] R.P. Lano and V.G.J. Rodgers, Mod. Phys. Lett. A7 (1992) 1725 \cr
\+ [8] R.P. Lano and V.G.J. Rodgers, UIOWA Preprint UIOWA-94-2,
hep-th/9401039\cr
\+~~~ ``A Study of Fermions Coupled to Gauge and Gravitational Fields on a
Cylinder'' \cr
\+ [9] R.J. Henderson and S.G. Rajeev, U Rochester Preprint
UR-1342,gr-qc/9401029 \cr
\+~~~  ``Quantum Gravity on a Circle and the Diffeomorphism Invariance\cr
\+~~~ of the Schroedinger Equation'' \cr
\+ [10] A. Einstein, Annalen der Physik, 49, 1919 \cr
\+ [11] L.D. Landau and E.M. Lifshitz, {\it The Classical Theory of Fields},\cr
\+(Fourth Revised English Edition, Pergamon Press, 1989) \cr

\end